\documentclass{article}
\usepackage{spconf,amsmath,graphicx}
\usepackage{booktabs,float,subfig}
\usepackage{CJKutf8}

\usepackage{cmap}
\title{CB-Conformer: Contextual biasing Conformer for biased word recognition}
\name{
\begin{tabular}{c}
Yaoxun Xu$^{1,*}$\thanks{* Work conducted when the first author was an intern at XVerse Inc.}, Baiji Liu$^2$, Qiaochu Huang$^1$, Xingchen Song$^1$,  \\\textit{Zhiyong Wu$^{1,3,4,\dagger}$\thanks{$\dagger$ Corresponding author.}, Shiyin Kang$^2$},\textit{ Helen Meng$^{4}$}
\end{tabular}
\vspace{-0.2cm}
}

\address{$^1$
    Shenzhen International Graduate School, Tsinghua University, Shenzhen, China \\
  $^2$ XVerse Inc., Shenzhen, China \quad 
    $^3$ Peng Cheng Lab, Shenzhen, China\\
    $^4$ The Chinese University of Hong Kong, Hong Kong SAR, China\\
    \small{
        \{xuyx22, hqc22, sxc19\}$@$mails.tsinghua.edu.cn, 
        zywu$@$sz.tsinghua.edu.cn,
        \{liubaiji,kangshiyin\}$@$xverse.cn
    }
}

\begin{document}
\ninept
%\linespread{0.92}\selectfont

%
\maketitle{}
\begin{abstract}
Due to the mismatch between the source and target domains, how to better utilize the biased word information to improve the performance of the automatic speech recognition model in the target domain becomes a hot research topic. Previous approaches either decode with a fixed external language model or introduce a sizeable biasing module, which leads to poor adaptability and slow inference. In this work, we propose CB-Conformer to improve biased word recognition by introducing the Contextual Biasing Module and the Self-Adaptive Language Model to vanilla Conformer. The Contextual Biasing Module combines audio fragments and contextual information, with only 0.2\% model parameters of the original Conformer. The Self-Adaptive Language Model modifies the internal weights of biased words based on their recall and precision, resulting in a greater focus on biased words and more successful integration with the automatic speech recognition model than the standard fixed language model. In addition, we construct and release an open-source Mandarin biased-word dataset based on WenetSpeech. Experiments indicate that our proposed method brings a 15.34\% character error rate reduction, a 14.13\% biased word recall increase, and a 6.80\% biased word F1-score increase compared with the base Conformer.

\end{abstract}
\section{Introduction}
\label{sec:intro}
End-to-end (E2E) automatic speech recognition (ASR) is gaining popularity due to its simple model structure, high training efficiency, and astounding performance across many tasks\cite{asr0,asr1}. In daily life, due to the diversity of scenarios, directly deploying an ASR model trained on one specific dataset to other domains will cause the problem of domain mismatch\cite{gourav2021personalization,mismatch0,mismatch2,mismatch3,mismatch100}.

There are unique words in specific domains, called biased words, such as Chinese person name \begin{CJK*}{UTF8}{gbsn}``曹操"
\end{CJK*} (CAO Cao in English). The domain mismatch is also reflected in the problem of biased word recognition. An intuitive and feasible way is introducing contextual information into the speech recognition process to emphasize specific biased words. 

There are two traditional approaches to integrating contextual information into speech recognition. The first approach is shallow fusion\cite{gourav2021personalization,lm0,lm1,lm2}, which fuses the task-specific external language model with the ASR model and boosts the scores of biased words during decoding. The second approach is integrating the contextual module with the ASR model\cite{nn0,nn1} in an all-neural network to incorporate contextual information, leveraging the powerful modeling expressiveness of E2E neural networks\cite{li2022recent}.

\begin{figure*}[!htb]
\centering
\subfloat[CB-Conformer]{
\includegraphics[height=60mm]{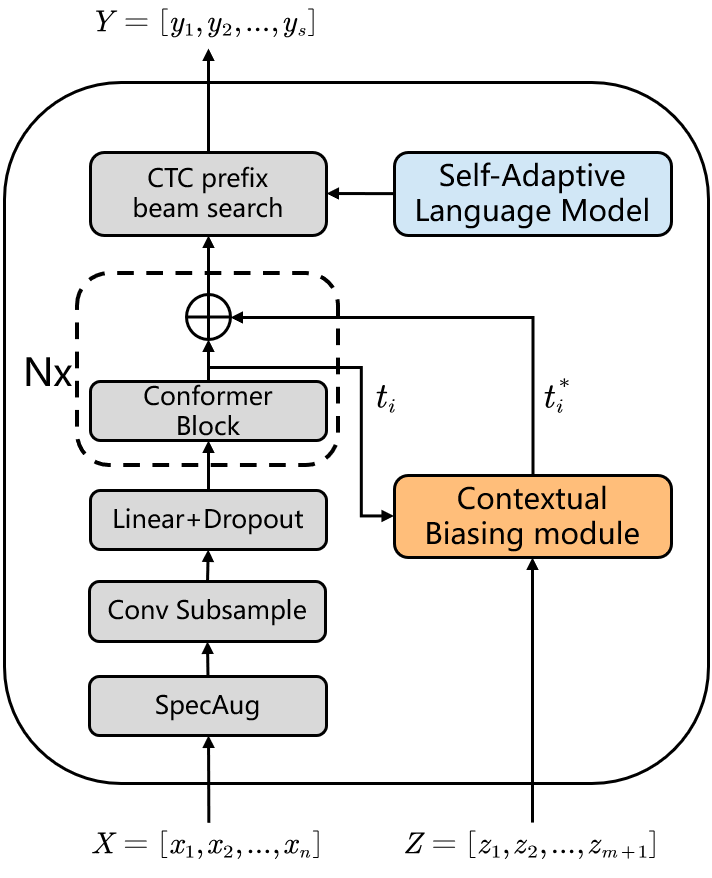}

\label{fig:lwunet}
}
\hspace{10mm}
\subfloat[Contextual Biasing Module]{
\includegraphics[height=60mm]{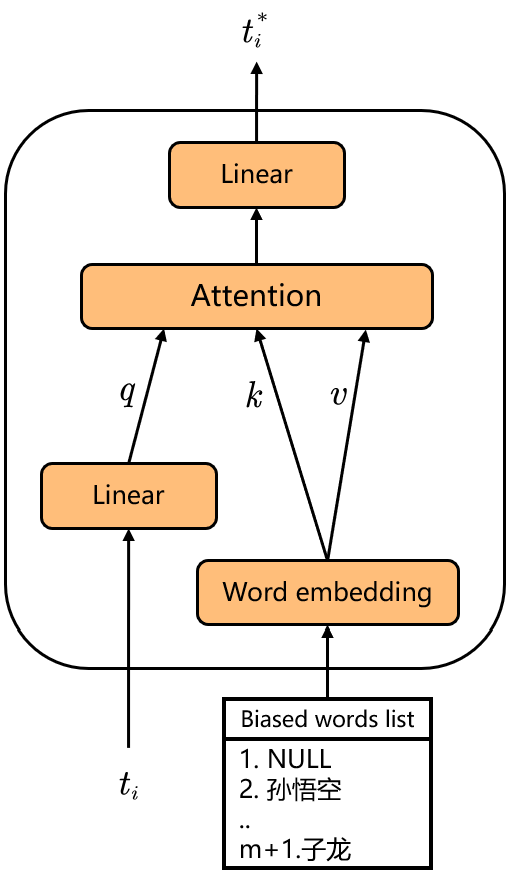}
\label{fig:sepresb}
}
\hspace{10mm}
\subfloat[Training Process of SA LM]{
\includegraphics[height=60mm]{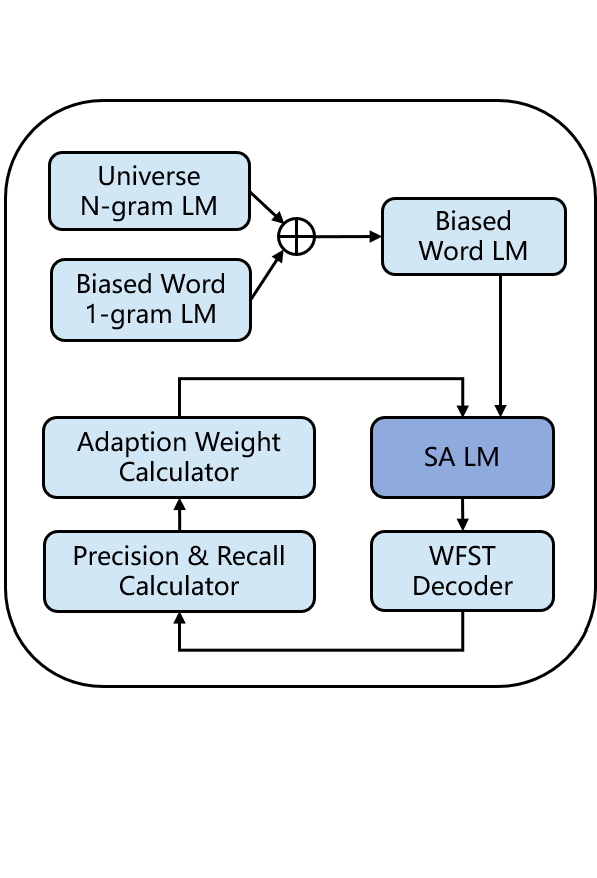}

\label{fig:lalayer}
}
\vspace{-0.2cm}
\caption{The overall architecture for CB-Conformer.}\
\label{fig:model_structure}
\vspace{-1cm}
\end{figure*}

However, the former approach suffers from the adaptation problem, where the traditional language model conflicts with the internal language model\cite{ilm,lm44} in the ASR model due to the inconsistency of the trained domains. In addition, the traditional language model maintains a consistent emphasis on biased words and is incapable of altering the weights of biased words, resulting in a poor prediction performance for biased words\cite{lm5,lm_33}. The latter approach employs an end-to-end model structure that co-trains the contextual module as part of the entire ASR model, making it impractical to regulate the degree of biasing in inference\cite{nn0,recent0,recent1}. Additionally, this approach tends to choose the recurrent neural network transducer (RNN-T)\cite{nn1,nn2} as the base model. However, the RNN-T model has a more complex structure and is more challenging to train, resulting in a decline in model prediction performance\cite{rnnt}. 

% but considering the multi-head attention mechanism of the Conformer model\cite{gulati2020conformer} greatly enhances the feature extraction capability of the model, the stronger feature extraction capability can better capture more relevant contextual information and increase the robustness of the model. At the same time, to keep the model concise and efficient, we inject contextual biasing into the Conformer-Encoder\cite{gulati2020conformer},  which required less dependency than the Decoder.

% In this work, a Contextual Biasing Module is proposed with a minimal number of parameters as part of the end-to-end speech recognition model, allowing contextual information to participate in the inference of the model. At the same time, an iteration algorithm based on shallow fusion is proposed by introducing the Self-Adaptive method to further improve the accuracy of biased word recognition and make for a better fit with ASR internal language model. We use the recall and precision of biased words as training loss to iteratively adjust the weight of words to increase the sensitivity to biased words.
In this work, we propose CB-Conformer to solve the problem of biased word recognition by augmenting the Conformer-Encoder\cite{gulati2020conformer} with a Contextual Biasing Module and a Self-Adaptive Language Model. The Self-Adaptive Language Model is proposed to increase the precision of biased word recognition and better coordinate with the ASR internal language model. To boost the sensitivity to biased words, we iteratively alter the weight of biased words in the Self-Adaptive Language Model depending on the recall and precision of biased words. The Contextual Biasing Module with an exceptionally minimal number of parameters is proposed as part of the E2E ASR model.
Due to its high applicability and extensibility, the Contextual Biasing Module can be attached to any block of the Conformer Encoder, and its training and decoding rates are fast. 
%Considering that the multi-head attention mechanism of the Conformer model significantly improves the model's ability to extract features, as well as its ease of training and small size, we select the Conformer as our base model.  

% In addition, there is no publicly available Mandarin dataset in the field of contextual biasing. We divide and filter the WenetSpeech dataset\cite{wenetspeech} and provide three specific types of datasets, which are the person name dataset, the place name dataset, and the organization name dataset. We also summarize a large contextual biasing dataset including the above three types, and all our experiments are based on the divided dataset.

In addition, as there is no open-source Mandarin biased words dataset, we construct and release the first open-source Mandarin biased words dataset. We provide three specific subsets filtered from the WenetSpeech\cite{wenetspeech} dataset: the person-name dataset, the place-name dataset, and the organization-name dataset. In addition, we combine the three sub-datasets into a composite biased words dataset and produce a dataset devoid of biased words. All our experiments are based on the divided datasets.

\section{Methods}
\label{sec:format}
% In general, two methods are proposed here. One is the Contextual Biasing Module for the Conformer model, and the other is the Self-Adaptive Language Model.
Fig.1(a) illustrates the overall architecture of our CB-Conformer for biased words recognition, which consists of a Contextual Biasing Module (shown in Fig.1(b)) and a Self-Adaptive Language Model (shown in Fig.1(c)). We choose the Conformer model for its strong global and local feature extraction ability and straightforward structure, which leads to fast training speed.

\subsection{Contextual Biasing Module}
\label{ssec:subhead}
As shown in Fig.1(b), a Contextual Biasing Module is introduced to extract the information from audio fragments and biased words using the attention mechanism\cite{vaswani2017attention}. The Contextual Biasing Module is relatively compact and primarily comprises a word embedding module and a contextual attention module.

We use the word embedding module to transform the biased words into feature vectors to have a more effective incorporation of audio information. To reduce the probability of overfitting and increase the generalizability of the ASR model, we add a ``NULL'' to the list of biased words to indicate that the current audio fragment is unrelated to any of the biased words. Thus, when given a set of biased words $Z=[z_1,z_2,... ,z_{m}]$, a total of m+1 words will be fed into the word embedding module and m+1 feature vectors $C=[c_1,c_2,...,c_{m+1}]$ are extracted. The contextual attention module aims to combine the current audio fragment with biased word information and identify the biased word that is most pertinent to the current audio fragment. Specifically, we use the output from a Conformer Block $t_i$ as the query vector followed by a linear layer, and the $c_j$ obtained from a certain biased word $z_j$ as the key and value vectors in the attention mechanism. The contextual attention module can identify the most appropriate words to the current audio fragment, and the output of this Conformer Block is updated by element-wise addition of the output of the Contextual Biasing Module $t_{i}^{*}$. We connect the output of each Conformer Block to the Contextual Biasing Module, allowing the contextual information to become more involved in the decoding process of audio. This connection also enables the ASR model to increase its sensitivity to biased words. 

\subsection{Self-Adaptive Language Model}
We also propose the Self-Adaptive (SA) language model (LM) to better extract the information of biased words from the standpoint of the text and to identify each biased word more precisely by modifying its weight in SA LM.
Given that the internal language model (ILM) implicated in the ASR model is trained in a different domain than the external language model (ELM), if the ELM connects directly to the ASR model, the prediction result $p(y|x)$ will conflict with the prediction result $p(y)$ of the ELM, which is equivalent to forecasting the result twice and will reduce the effectiveness of the ELM. However, SA LM improves the ASR model's recognition performance by adjusting the biased words' weights to be more compatible with the ASR ILM, minimizing the likelihood of conflicts with the ASR ILM and focusing more on the biased words.  

As illustrated in Fig.1(c), The SA LM is constructed by combining the N-gram LM (trained with universe text) and the 1-gram LM (trained with biased words). The SA LM is represented as WFST\cite{mohri2002weighted} and connects to the ASR model for decoding using shallow fusion. 
During training, the probability of biased words and their backoff probability can be adjusted by calculating the precision and recall in the sentence derived from decoding, resulting in better biased word recognition than normal LM. Specifically, the iterative formula for the biased weights is as follows:
\begin{equation}
\setlength{\abovedisplayskip}{3pt}
\label{equ:1}
    lr_i= \eta * lr_{i-1}
\end{equation}
\begin{equation}
\label{equ:2}
    w_{k,i} = w_{k,i-1} + sgn(\alpha_k - \beta_k) * lr_{i} * \Delta   
\end{equation}

Where $lr_i$ is the learning rate of iteration $i$, $\eta$ is the decay rate of the learning rate for each iteration, $w_{k,i}$ is the biased weight for $k_{th}$ biased word, $sgn$ is the sign function, $\alpha_k$ is the precision of $k_{th}$ biased word, $\beta_k$ is the recall of $k_{th}$ biased word, and $\Delta$ is the weight step. This function will loop until the precision is close to the recall. Optionally, $\beta_k$ can be set as a constant number to precisely control the precision of the ASR model. Finally, the new weight of each word will be updated to the corresponding log-scale weight in the LM model after each iteration.

\subsection{Calculation of biased word recall and precision}
This method uses the shortest edit-distance path to consider the mismatch of the biased word's position. The mismatch cannot be disregarded if the sequence is too long and contains numerous biased words. When dealing with Chinese sequence, the following process is followed: 

1. Tokenize the sequence in a unique way.
Once a biased word is identified in the sequence, it will be treated as a separate unit, while all other units will be considered characters.
If a prospective shorter biased word is the prefix of a longer biased word, only the longer biased word is examined. 

2. Generate the edit-distance alignment (the path with minimal edit-distance) for the tokenized sequence of biased words.
The edit distance of the biased word unit position is 0 if the matching alignment position is identical; otherwise, it is 1.
Optionally, if two separate biased words share the same meaning, an additional mapping post-processing will be included to maintain their edit distance as 0.

3. Scan the alignment again, and statistics the biased word unit position, then precision and recall are calculated as below:
\begin{equation}
\setlength{\abovedisplayskip}{3pt}
\label{equ:1}
    precision = M/R
\end{equation}
\begin{equation}
\label{equ:2}
    recall = M/L  
\end{equation}
Where $M$ is the number of biased words that simultaneously appear in the relevant positions in the label and result sequence, $L$ is the number of the biased words appearing in the label sequence, and $R$ is the number of the biased words appearing in the result sequence.

\label{ssec:subhead}

\section{EXPERIMENTS}
\label{sec:pagestyle}
\vspace{-0.2cm}
\subsection{Datasets}

Since there is no open-source Mandarin dataset for biased words, we divide and publish the Mandarin biased words dataset based on the 1000-hour WenetSpeech dataset. In particular, we split sub-datasets into three categories: person-name dataset, place-name dataset, and organization-name dataset. First, we use the hanlp\footnote{https://github.com/hankcs/HanLP} tool to identify the named entities for the entire 1000-hour dataset, and then we categorize the named entities. Using the person-name dataset as an illustration, we obtain a list of biased words for Chinese person names. For each biased word, we search the original dataset for all sentences containing the word and retain the words in the three biased word lists with frequencies between 5 and 700. The biased word dataset is divided into train, test, and dev sets.
\begin{table}[h]
\label{label1}
\centering
\caption{Details of Proposed Dataset, containing three sub-datasets, a total biased dataset comprising the aforementioned sub-datasets, and a base dataset devoid of biased words. }
\setlength{\tabcolsep}{0.7mm}
\scalebox{0.9}{
\begin{tabular}{cccccc}
\toprule[2pt]
             & \multicolumn{4}{c}{dataset}           & number of \\ \cline{2-5}
             & train    & dev  & test   & time\_last (h) & biased words           \\ \hline
person-name       & 1000     & 1000 & 9997   & 10.24      & 73         \\
place-name        & 1000     & 1000 & 10191  & 12.96      & 42         \\
organization-name & 1000     & 500 & 2035  & 3.94      & 183       \\
all\_biased  & 2992     & 2497 & 21972 & 26.77     & 298       \\
no\_biased   & 12774458 & /    & /      & 793.30     & /                             \\
\bottomrule[2pt]
\end{tabular}
}
\end{table}
Since the biased words are uncommon and the corresponding number of sentences is relatively small, we set the size of the train set for each specific dataset to 1000 utterances, which is much smaller than the size of the test set, to accommodate the usage scenario for biased words better. As summarized in Table 1, we also generate a total biased words dataset called ``all\_biased'' and a dataset without biased words called ``no\_biased'', and all of our experiments are based on these datasets. Instructions and 
datasets are available at
 GitHub repo\footnote{ https://github.com/thuhcsi/Contextual-Biasing-Dataset/ }.

\subsection{Experimental Setup}
Generally, we keep the training procedure the same as described in  \cite{yao2021wenet}.
% 80-dimension log Mel-filter banks(FBANK) are computed on a 25 milliseconds window, with 10 milliseconds overlap between neighboring windows. We use a 5149-dimension character-based dictionary generated from training data for text tokenization. For the selection of parameters of the model, we use an Encoder consisting of 12 layers Conformer Blocks. The Conformer block consists of a feed-forward layer with 2048 nodes, a transformer layer with 4 256-dim attention heads, a convolutional module with kernel size=15, and the last feed-forward layer with 2048 nodes. SiLU\cite{silu} is used as the activation function and dropout is set to 0.1 when training. For the prediction part, we choose not to use Decoder but only CTC prefix beam search as decoding, the outputs from the encoder are projected through a feed-forward layer to 5149, which is the size of the dictionary, and use CTC Loss as the loss function.
For each of our experiments, we only use CTC-loss as the loss function to increase training and decoding speed. The only component of our base model is the Encoder module, which consists of 12 Conformer Blocks, and no Decoder module is included. We use a 5149-dimension character-based dictionary generated from training data for text tokenization. Apart from that, we keep the same training and testing settings as in open-sourced WeNet recipes\footnote{https://github.com/wenet-e2e/wenet/tree/main/examples/wenetspeech/s0}, including model regularization (weight decay, etc.), optimizer, learning rate schedule, data augmentation, etc.

The Word Embedding module converts a biased word $z_j$ to a 64-dimension contextual embedding $c_j$ before passing it to the Contextual attention module.
A 64-node feed-forward layer is used to project the 256-dimension output of a Conformer Block $t_i$ to 64-dimension $t_{i}^{'}$, which are computed by the attention mechanism with $c_j$ and projected to 256 dimensions by the linear layer to get $t_{i}^{*}$, which is the same dimension as $t_i$. In this setting, the Contextual Biasing Module total makes up \textless 40k parameters, accounting for less than 0.2\% of the base model parameters. In SA LM, we set $lr_0$=1, $\eta$=0.9, $\Delta$=1, $\beta$=0.98.

Our evaluation metrics are character error rate (CER), biased word recall, biased word precision, and F1-score.
\vspace{-0.3cm}
\section{RESULTS}
\label{sec:typestyle}
The outcomes presented in Table 2 and Table 4 are all evaluated using the person-name dataset. S1 trains the vanilla Conformer model on the no\_biased dataset from scratch. S2 is acquired via shallow fusion of S1 and the LM trained by the person-name train set, whereas S3 is obtained by finetuning the model on the person-name train set based on S1 without modifying the model structure. 

\begin{table}[h]
\centering
\caption{Comparisons of the effectiveness of the Self-Adaptive Language Model (SA LM) and the Contextual Biasing Module (CBM) on the person-name dataset (\%).}
\setlength{\tabcolsep}{2.8mm}
\begin{tabular}{lcccc}
\toprule[2pt]
                     & CER   & Recall & Precision & F1    \\ \hline
S1: baseline          & 18.28 & 0.277  & \textbf{0.997}    & 0.434 \\
S2: S1+LM             & 13.97 & 0.654  & 0.991    & 0.788 \\
S3: S1+finetune       & 10.56 & 0.821  & 0.988    & 0.897 \\
S4: S1+CBM     & 10.68 & 0.897  & 0.987    & 0.940 \\
S5: S3+LM            & 9.91 & 0.904 & 0.986 & 0.943\\
S6: S1+SA LM          & 11.97 & 0.867  & 0.924    & 0.895 \\
S7 :S3+SA LM          & 9.24  & 0.925  & 0.981    & 0.952 \\
S8 :S4+SA LM          & \textbf{8.94}  & \textbf{0.937}  & 0.980    & \textbf{0.958} \\
\bottomrule[2pt]
\end{tabular}
\end{table}

\begin{table*}[ht]
\caption{Results on the place-name dataset and the organization-name dataset (\%).}
\centering
\setlength{\tabcolsep}{3.8mm}
\scalebox{1.0}{
\begin{tabular}{lcccclcccc}
\toprule[2pt]
                     & \multicolumn{4}{c}{place-name dataset} &  & \multicolumn{4}{c}{organization-name dataset} \\ \cline{2-5} \cline{7-10} 
\multicolumn{1}{c}{} & CER    & Recall & Precision & F1 &  & CER      & Recall    & Precision    & F1      \\ \hline
S1: baseline         & 16.14  & 0.259  & 0.999     & 0.411    &  & 12.01    & 0.466     & 0.99         & 0.634   \\
S9: baseline+finetune & 9.86   & 0.884  & 0.996     & 0.937    &  & 7.17     & 0.849     & 0.999        & 0.918   \\
S10: baseline+CBM & 10.11  & 0.931  & 0.995     & 0.962    &  & 7.16     & 0.896     & 0.999        & 0.945  \\
\bottomrule[2pt]
\end{tabular}
}
\vspace{-0.3cm}
\end{table*}

As demonstrated in Table 2, S1 (baseline) performs poorly on the person-name dataset, with a recall of 0.277 and an F1-score of 0.435.
Due to the fusion of the traditional language model, S2 has a 23.58\% lower CER and a 136.10\% higher recall than S1.
Due to finetuning, S3 has a 42.23\% lower CER and a 106.68\% higher F1 score than S1. 

Compared to S1, S4 introduces the Contextual Biasing Module.
S4 has a 9.26\%/4.79\% increase in recall and F1-score, respectively, as compared to S3, as a result of the feature extraction of biased words and the combination of contextual features and audio features. 
Based on the foundations established in S1, S3, and S4, we augmented S6, S7, and S8 with SA LM trained on the person-name dataset. All the findings obtained after fusing SA LM are superior to those obtained without fusion, indicating the validity and consistency of SA LM. 
Due to the SA LM, S6 is more sensitive to biased words than S2, with a 14.32\% lower CER and a 13.58\% higher F1-score. Additionally, S7 has improved recognition of biased words compared to S5, with a 6.76\% lower CER and a 0.95\% higher F1-score. These results demonstrate that the SA LM is effective in various asr model structures, with more sensitivity and adaptability to biased words than the standard LM.
Comparing S7 and S8 to S3 and S4, it is apparent that SA LM enhances the recognition accuracy of the ASR model with or without the Contextual Biasing Module, resulting in lower CER/higher F1-score of 12.5\%/6.13\%, 16.29\%/1.91\%, respectively. 
When the Contextual Biasing Module incorporates with the SA LM, the proposed model (S8) reaches the lowest CER and the highest F1-score, achieving 8.94\%/0.958 in the test set of the person-name dataset. 
\vspace{-0.2cm}
\subsection{The generalizability of Contextual Biasing Module}

The Contextual Biasing Module also achieves excellent results concerning the place-name and organization-name datasets.

As demonstrated in Table 3, comparing S10 (baseline model with Contextual Biasing Module) to S9 (baseline finetune on the specific dataset), the increase of recall/F1-score for the place-name dataset and organization-name dataset are as follows: 5.32\%/2.67\%,    5.54\%/2.94\%, respectively. The generalizability of the Contextual Biasing Module is demonstrated by the fact that our proposed Contextual Biasing Module shows noticeable advancements in biased word recognition across multiple datasets.

\vspace{-0.2cm}
\subsection{The requirement for freezing base model parameters }

We also find it essential to freeze the parameters of the base model (Conformer) when training the Contextual Biasing Module.

\begin{table}[h]
\centering
\caption{Results on whether frozen the Conformer in the person-name dataset (\%)}
\setlength{\tabcolsep}{1.5mm}
\scalebox{1.1}{
\begin{tabular}{lcccc}
\toprule[2pt]
                          & CER   & Recall & Precision & F1    \\ \hline
S3: baseline+finetune      & 10.56 & 0.821  & 0.988    & 0.897 \\
S4: CBM\_freeze     & 10.68 & 0.897  & 0.987    & 0.940 \\
S11: CBM\_no\_freeze & 16.82 & 0.849  & 0.938    & 0.891 \\
\bottomrule[2pt]
\end{tabular}
}
\end{table}
Table 4 reveals that the CER of S11 (without freezing the parameters of the Conformer) is 59.28\% higher than that of S3 in the person-name dataset.
In comparison, the CER of S4 (freezing parameters of the Conformer) is marginally 1.14\% higher than that of S1. We hypothesize that the poor performance without freezing parameters is that the person-name train set consists of just 1,000 utterances, causing the overfitting of the model due to overtraining. Whether the parameters of the Conformer are frozen or not, adding the Contextual Biasing Module increases the recall of biased words by 9.26\%/3.416\%, respectively, reflecting the Contextual Biasing Module's ability to capture the correlation between biased words and audio fragments.

% \subsection{SA LM Weights Analysis}

\begin{figure}[ht]

\centering
\subfloat[The variation of the weights of biased words in SA LM]{
% \begin{minipage}{6cm}
\centering
\includegraphics[width=0.9\linewidth]{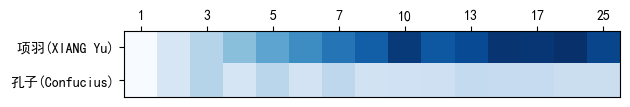}
% \end{minipage}
}

\subfloat[The variation of recall of biased words]{
% \begin{minipage}{6cm}
\centering
\includegraphics[width=0.9\linewidth]{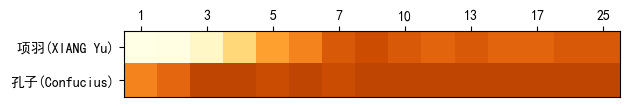}
% \end{minipage}
}
\caption{Typical cases for the changes in biased words' weights and recall in the person-name test set: the number of training iterations is listed on the top, while some biased words (Chinese person names) are listed on the left. Larger weights have a darker color.}
\vspace{-0.5cm}
\end{figure}

\subsection{Weight changes for biased words in SA LM}
As depicted in Fig.2, the two words' weights in SA LM are identical before training, and after training, \begin{CJK*}{UTF8}{gbsn}
``项羽
\end{CJK*}'' (XIANG Yu in English) has a larger weight than \begin{CJK*}{UTF8}{gbsn}
``孔子
\end{CJK*}'' (Confucius in English). Before incorporating SA LM (S3), the biased word ``Confucius'' has a higher recall in the person test set than ``XIANG Yu''. 
As the number of training iterations increases, the recall on the test set for both ``XIANG Yu'' and ``Confucius'' increases. However, the weight of ``XIANG Yu'' in SA LM gradually increases, whereas ``Confucius'' shows no significant difference. We hypothesize that the ILM in the ASR model has already learned some biased words, such as ``Confucius'', so that after the addition of SA LM, SA LM will not pay too much attention to these biased words and their relative weights in SA LM will not change significantly. SA LM will pay more attention to words not previously learned by the ILM, such as ``XIANG Yu''. The weights of these biased words will change more significantly during training, reflecting that SA LM can better cooperate with the ILM of ASR to reduce the possibility of conflicts and increase the performance of biased word recognition.

% We conjecture that before combining SA LM with the ASR model, the ILM of ASR model acquires knowledge about ``Confucius'' through training, but does not acquire knowledge about ``XIANG Yu''. Therefore, when combining SA LM with the ASR model, SA LM tends to focus more on ``XIANG Yu'', i.e., increase the weight of ``XIANG Yu'' in the SA LM, while SA LM does not focus excessively on ``Confucius'' during training because the ILM in the ASR model has acquired some knowledge about ``Confucius'', and the weight of ``Confucius'' in SA LM will be lower and more stable. This also demonstrates the capacity of SA LM to work more effectively with ASR's ILM to improve the recognition of biased words. 
\vspace{-0.2cm}
\section{CONCLUSION}
\label{sec:majhead}

To improve the recognition of biased words, we propose CB-Conformer, a novel technique that introduces the Contextual Biasing Module and the Self-Adaptive Language Model to the original Conformer model.
We also construct and release an open-source Mandarin biased words dataset.
The tiny Contextual Biasing Module incorporates biased word information into the inference process as part of the E2E ASR model.
The Self-Adaptive Language Model updates the weights of biased words based on their respective recall and precision, enhancing its compatibility with the ILM of the ASR model and its capability of capturing contextual information.
Evaluated on the constructed dataset, CB-Conformer has the lowest CER and the highest F1-score for biased word training with only 1000 utterances. The results also indicate that CB-Conformer has better mobility and stability with its intense sensitivity to biased words.

\textbf{Acknowledgement}: This work is supported by Shenzhen Science and Technology Program (WDZC20200818121348001), the Major Key Project of PCL (PCL2021A06, PCL2022D01) and AMiner.Shenzhen SciBrain fund.

\bibliographystyle{IEEEbib}
\bibliography{strings,refs}

\begin{thebibliography}{10}

\bibitem{asr0}
Yanzhang He, Tara~N Sainath, Rohit Prabhavalkar, Ian McGraw, Raziel Alvarez,
  Ding Zhao, David Rybach, Anjuli Kannan, Yonghui Wu, Ruoming Pang, et~al.,
\newblock ``Streaming end-to-end speech recognition for mobile devices,''
\newblock in {\em ICASSP}. IEEE, 2019, pp. 6381--6385.

\bibitem{asr1}
Bo~Li, Shuo-yiin Chang, Tara~N Sainath, Ruoming Pang, Yanzhang He, Trevor
  Strohman, and Yonghui Wu,
\newblock ``Towards fast and accurate streaming end-to-end asr,''
\newblock in {\em ICASSP}. IEEE, 2020, pp. 6069--6073.

\bibitem{gourav2021personalization}
Aditya Gourav, Linda Liu, Ankur Gandhe, Yile Gu, Guitang Lan, Xiangyang Huang,
  Shashank Kalmane, Gautam Tiwari, Denis Filimonov, Ariya Rastrow, et~al.,
\newblock ``Personalization strategies for end-to-end speech recognition
  systems,''
\newblock in {\em ICASSP}. IEEE, 2021, pp. 7348--7352.

\bibitem{mismatch0}
Wei-Ning Hsu, Yu~Zhang, and James Glass,
\newblock ``Unsupervised domain adaptation for robust speech recognition via
  variational autoencoder-based data augmentation,''
\newblock in {\em ASRU}. IEEE, 2017, pp. 16--23.

\bibitem{mismatch2}
Zhong Meng, Sarangarajan Parthasarathy, Eric Sun, Yashesh Gaur, Naoyuki Kanda,
  Liang Lu, Xie Chen, Rui Zhao, Jinyu Li, and Yifan Gong,
\newblock ``Internal language model estimation for domain-adaptive end-to-end
  speech recognition,''
\newblock in {\em SLT}. IEEE, 2021, pp. 243--250.

\bibitem{mismatch3}
Khe~Chai Sim, Fran{\c{c}}oise Beaufays, Arnaud Benard, Dhruv Guliani, Andreas
  Kabel, Nikhil Khare, Tamar Lucassen, Petr Zadrazil, Harry Zhang, Leif
  Johnson, et~al.,
\newblock ``Personalization of end-to-end speech recognition on mobile devices
  for named entities,''
\newblock in {\em ASRU}. IEEE, 2019, pp. 23--30.

\bibitem{mismatch100}
Wei-Ning Hsu, Yu~Zhang, and James Glass,
\newblock ``Unsupervised domain adaptation for robust speech recognition via
  variational autoencoder-based data augmentation,''
\newblock in {\em 2017 IEEE Automatic Speech Recognition and Understanding
  Workshop (ASRU)}. IEEE, 2017, pp. 16--23.

\bibitem{lm0}
Anjuli Kannan, Yonghui Wu, Patrick Nguyen, Tara~N Sainath, Zhijeng Chen, and
  Rohit Prabhavalkar,
\newblock ``An analysis of incorporating an external language model into a
  sequence-to-sequence model,''
\newblock in {\em ICASSP}. IEEE, 2018, pp. 1--5828.

\bibitem{lm1}
Ding Zhao, Tara~N Sainath, David Rybach, Pat Rondon, Deepti Bhatia, Bo~Li, and
  Ruoming Pang,
\newblock ``Shallow-fusion end-to-end contextual biasing.,''
\newblock in {\em Interspeech}, 2019, pp. 1418--1422.

\bibitem{lm2}
Duc Le, Gil Keren, Julian Chan, Jay Mahadeokar, Christian Fuegen, and Michael~L
  Seltzer,
\newblock ``Deep shallow fusion for rnn-t personalization,''
\newblock in {\em SLT}. IEEE, 2021, pp. 251--257.

\bibitem{nn0}
Minglun Han, Linhao Dong, Zhenlin Liang, Meng Cai, Shiyu Zhou, Zejun Ma, and
  Bo~Xu,
\newblock ``Improving end-to-end contextual speech recognition with
  fine-grained contextual knowledge selection,''
\newblock in {\em ICASSP}. IEEE, 2022, pp. 8532--8536.

\bibitem{nn1}
Tsendsuren Munkhdalai, Khe~Chai Sim, Angad Chandorkar, Fan Gao, Mason Chua,
  Trevor Strohman, and Fran{\c{c}}oise Beaufays,
\newblock ``Fast contextual adaptation with neural associative memory for
  on-device personalized speech recognition,''
\newblock in {\em ICASSP}. IEEE, 2022, pp. 6632--6636.

\bibitem{li2022recent}
Jinyu Li et~al.,
\newblock ``Recent advances in end-to-end automatic speech recognition,''
\newblock {\em APSIPA Transactions on Signal and Information Processing}, vol.
  11, no. 1, 2022.

\bibitem{ilm}
Mohammad Zeineldeen, Aleksandr Glushko, Wilfried Michel, Albert Zeyer, Ralf
  Schl{\"u}ter, and Hermann Ney,
\newblock ``Investigating methods to improve language model integration for
  attention-based encoder-decoder asr models,''
\newblock in {\em Interspeech}, 2021, pp. 2856--2860.

\bibitem{lm44}
Janne Pylkk{\"o}nen, Antti Ukkonen, Juho Kilpikoski, Samu Tamminen, and Hannes
  Heikinheimo,
\newblock ``Fast text-only domain adaptation of rnn-transducer prediction
  network,''
\newblock in {\em Interspeech}, 2021, pp. 1882--1886.

\bibitem{lm5}
Wei Zhou, Zuoyun Zheng, Ralf Schl{\"u}ter, and Hermann Ney,
\newblock ``On language model integration for rnn transducer based speech
  recognition,''
\newblock in {\em ICASSP}. IEEE, 2022, pp. 8407--8411.

\bibitem{lm_33}
Mohammad Zeineldeen, Aleksandr Glushko, Wilfried Michel, Albert Zeyer, Ralf
  Schl{\"u}ter, and Hermann Ney,
\newblock ``Investigating methods to improve language model integration for
  attention-based encoder-decoder asr models,''
\newblock in {\em Interspeech}, 2021, pp. 2856--2820.

\bibitem{recent0}
Dhruv Jain, Khoa Huynh Anh~Nguyen, Steven M.~Goodman, Rachel Grossman-Kahn,
  Hung Ngo, Aditya Kusupati, Ruofei Du, Alex Olwal, Leah Findlater, and Jon
  E.~Froehlich,
\newblock ``Protosound: A personalized and scalable sound recognition system
  for deaf and hard-of-hearing users,''
\newblock in {\em CHI Conference on Human Factors in Computing Systems}, 2022,
  pp. 1--16.

\bibitem{recent1}
Vrunda~N Sukhadia and S~Umesh,
\newblock ``Domain adaptation of low-resource target-domain models using
  well-trained asr conformer models,''
\newblock {\em arXiv preprint arXiv:2202.09167}, 2022.

\bibitem{nn2}
Kanthashree~Mysore Sathyendra, Thejaswi Muniyappa, Feng-Ju Chang, Jing Liu,
  Jinru Su, Grant~P Strimel, Athanasios Mouchtaris, and Siegfried Kunzmann,
\newblock ``Contextual adapters for personalized speech recognition in neural
  transducers,''
\newblock in {\em ICASSP}. IEEE, 2022, pp. 8537--8541.

\bibitem{rnnt}
Anshuman Tripathi, Han Lu, Hasim Sak, and Hagen Soltau,
\newblock ``Monotonic recurrent neural network transducer and decoding
  strategies,''
\newblock in {\em ASRU}. IEEE, 2019, pp. 944--948.

\bibitem{gulati2020conformer}
Anmol Gulati, James Qin, Chung-Cheng Chiu, Niki Parmar, Yu~Zhang, Jiahui Yu,
  Wei Han, Shibo Wang, Zhengdong Zhang, Yonghui Wu, et~al.,
\newblock ``Conformer: Convolution-augmented transformer for speech
  recognition,''
\newblock in {\em Interspeech}, 2020, pp. 5036--5040.

\bibitem{wenetspeech}
Binbin Zhang, Hang Lv, Pengcheng Guo, Qijie Shao, Chao Yang, Lei Xie, Xin Xu,
  Hui Bu, Xiaoyu Chen, Chenchen Zeng, et~al.,
\newblock ``Wenetspeech: A 10000+ hours multi-domain mandarin corpus for speech
  recognition,''
\newblock in {\em ICASSP}. IEEE, 2022, pp. 6182--6186.

\bibitem{vaswani2017attention}
Ashish Vaswani, Noam Shazeer, Niki Parmar, Jakob Uszkoreit, Llion Jones,
  Aidan~N Gomez, {\L}ukasz Kaiser, and Illia Polosukhin,
\newblock ``Attention is all you need,''
\newblock {\em Advances in neural information processing systems}, vol. 30,
  2017.

\bibitem{mohri2002weighted}
Mehryar Mohri, Fernando Pereira, and Michael Riley,
\newblock ``Weighted finite-state transducers in speech recognition,''
\newblock {\em Computer Speech \& Language}, vol. 16, no. 1, pp. 69--88, 2002.

\bibitem{yao2021wenet}
Zhuoyuan Yao, Di~Wu, Xiong Wang, Binbin Zhang, Fan Yu, Chao Yang, Zhendong
  Peng, Xiaoyu Chen, Lei Xie, and Xin Lei,
\newblock ``Wenet: Production oriented streaming and non-streaming end-to-end
  speech recognition toolkit,''
\newblock in {\em Interspeech}, 2021, pp. 4054--4058.

\end{thebibliography}

\end{document}